\documentclass[prb,aps,superscriptaddress,twocolumn]{revtex4}
\usepackage{graphicx}
\usepackage{dcolumn}
\usepackage{bm}
\usepackage{amssymb,ulem}
\usepackage[usenames]{color}

\begin{document}


\title{Raman scattering in a Heisenberg {\boldmath $S=1/2$} antiferromagnet on the anisotropic  triangular
lattice}

\author{Natalia B. Perkins}
\affiliation{Department of Physics, University of Wisconsin-Madison,
Madison, WI 53706, USA}

\author{Gia-Wei Chern}
\affiliation{Department of Physics, University of Wisconsin-Madison,
Madison, WI 53706, USA}
\affiliation{Theoretical Division, Los Alamos National Laboratory, Los Alamos, New Mexico 87545, USA}

\author{Wolfram Brenig}
\affiliation{Institute for Theoretical Physics, Technical University
Braunschweig, Mendelssohnstr. 3, 38106 Braunschweig, Germany}

\begin{abstract}
We investigate the two-magnon Raman scattering from an anisotropic $S=1/2$ triangular Heisenberg antiferromagnet Cs$_2$CuCl$_4$.
We find that the Raman response is very sensitive to  magnon-magnon
interactions and to scattering geometries, a feature that is in remarkable contrast with the polarization-independent
Raman signal from the isotropic triangular Heisenberg antiferromagnet.
Since a spin-liquid ground state gives rise to a similar rotationally invariant Raman response,
our results on the polarization dependence of the scattering spectrum suggest that Raman spectroscopy provides
a useful probe, complementary to neutron scattering, of the ground-state properties of Cs$_2$CuCl$_4$,
particularly whether the time-reversal symmetry is broken in the ground state.
\end{abstract}

\date{\today}

\maketitle

\section{Introduction}
Recently Heisenberg antiferromagnets on the triangular lattice have attracted
considerable experimental and theoretical interest. Among them, Cs$_2$CuCl$_4$ has
  been under particular scrutiny as it provides an interesting example of a
spatially anisotropic spin-$1/2$ triangular
antiferromagnet. \cite{coldea1,coldea2,
  isakov05,alicea05,yunoki06,veillette05,dalidovich06,starykh07,starykh10,kresel11}
Much of the interest
in this compound stems from its unusual,
non-classical magnetic properties, arising from  the competition between the spatial anisotropy, Dzyaloshinskii-Moriya (DM) interactions, and quantum
fluctuations.

Extensive neutron scattering studies \cite{coldea1,coldea2} on the magnetic properties of Cs$_2$CuCl$_4$
revealed several interesting features.
First,  despite frustration and low-dimensionality, a long-range magnetic order develops at low temperatures:
the observed spin order is incommensurate and sets in at
temperatures below $T_N=0.62$ K.  Magnetic excitations above this ground
state are also quite unusual.  While the low-energy excitation spectrum contains
well-defined sharp modes, as expected for an ordered state, a broad
continuum is formed at intermediate and high energies. A number of
theoretical proposals have been made
to explain the origin of this continuum. It has been suggested that the existence of a continuum is an
indication that the system is proximate to a spin liquid phase
that determines the behavior of excitations except for low energies.\cite{isakov05,alicea05,yunoki06,starykh07}
An alternative suggestion is that the continuum might originate from
magnon-magnon scattering which is enhanced in non-colinear
magnets.\cite{veillette05,dalidovich06}

In view of the ambiguity in the interpretation of the neutron scattering data, a complementary
experimental analysis of the magnetic properties of Cs$_2$CuCl$_4$ by a different technique
is highly desirable.  A very effective and frequently used experimental tool to study low-temperature
properties of low-dimensional quantum magnets is the two-magnon Raman
scattering.\cite{gozar05,gingras06,choi09,iliev10,wang11,chen11}
The two-magnon Raman intensity is directly related to the spectrum of two
interacting magnons in a total spin zero state at vanishingly small momentum and weighted by a form factor
that is dependent on the polarization of the incident light. It contains detailed
information on the two-magnon density of states and the magnon-magnon interactions.
Therefore, direct comparison of experimental spectra with those obtained from theoretical analysis can lead
to rather accurate estimates on  values of the superexchange and DM interactions.  In addition,
the analysis of the polarization dependence of the magnetic Raman scattering\cite{perkins08,cepas08} might
shed some light on whether the ground state is ordered, as neutron scattering experiments have suggested, or is
actually in a spin liquid state, as some theories suggest. A pronounced
polarization dependence would indicate that magnetically ordered state is the most probable candidate  for the ground state, and   that the observed continuum is due to
relatively strong interactions between magnons.
On the other hand, if the Raman scattering depends weakly on
the scattering geometry, the continuum in neutron scattering might be
due to unconventional excitations above a spin liquid ground state.

\begin{figure}
\includegraphics[width=0.85\columnwidth]{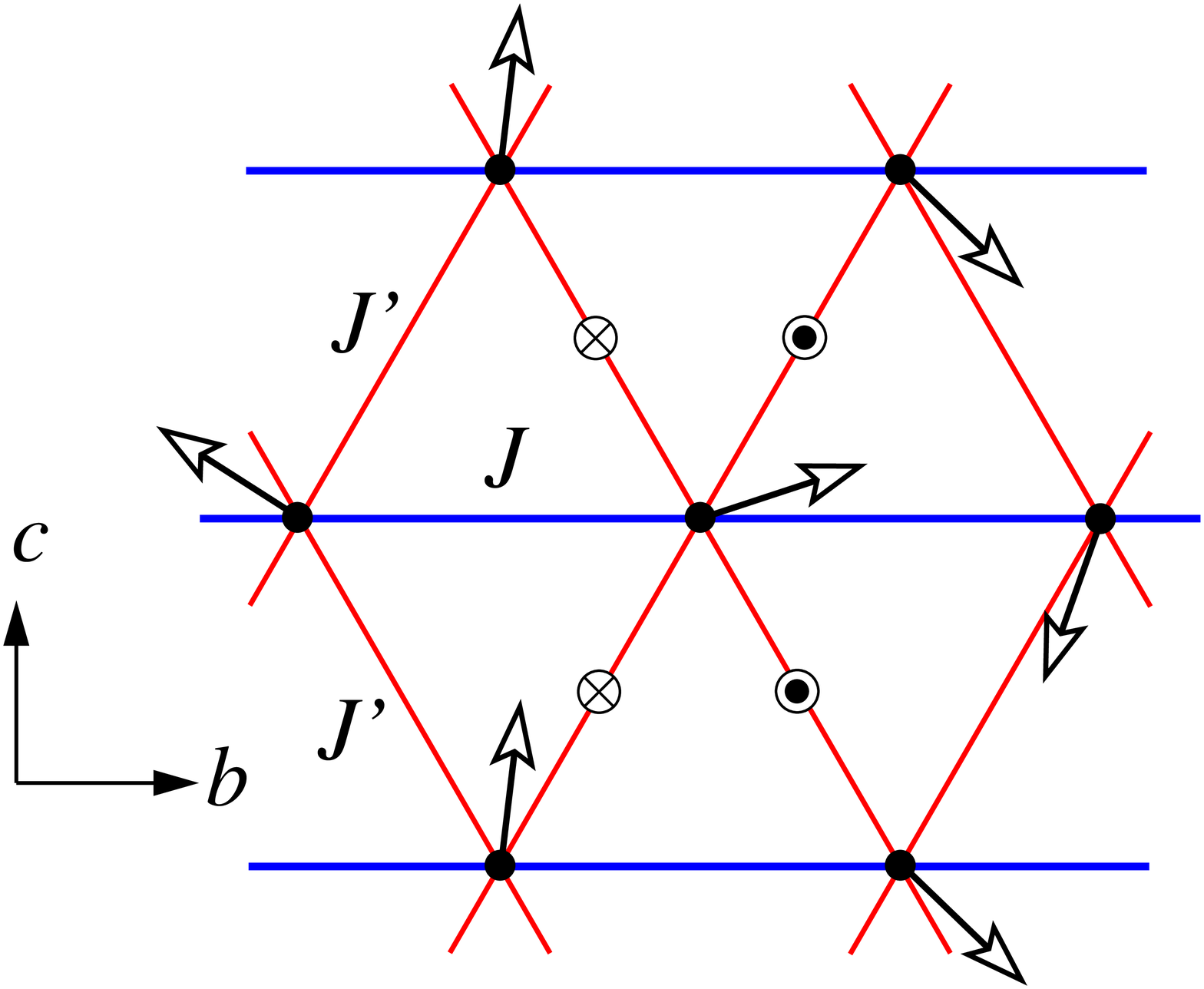}
\caption{\label{fig:tri} Triangular lattice with anisotropic spin
exchanges.}
\end{figure}

In this paper, we
carry out a theoretical analysis of two-magnon Raman
scattering from an anisotropic $S=1/2$ Heisenberg antiferromagnet (HAF) on a
triangular lattice (see Fig.\ref{fig:tri}).
To evaluate the two-magnon Raman spectra, we use the well-established, semi-phenomenological Loudon-Fleury (LF)
approach.\cite{fleury}
By chosing model parameters relevant for the compound Cs$_2$CuCl$_4$,
we find that the spectral shape as a function of frequency
is sensitive to $1/S$ corrections of the magnon spectrum and is strongly
modified by the magnon-magnon interactions in the final state. The intensity of the
two-magnon peak also significantly depends on the scattering geometry which is
in  contrast with the polarization independence of the magnetic Raman response
in isotropic HAF on triangular lattice.\cite{perkins08}

The paper is organized as follows. In Sec.~\ref{sec:model}, we present a
model of Cs$_2$CuCl$_4$ and discuss its classical ground state, which is an
incommensurate spin spiral with a pitch vector determined by the competition
of anisotropic nearest-neighbor interactions. In Sec.~\ref{sec:largeS}, we first
review results of the one magnon excitations in the anisotropic $S=1/2$ HAF to
first order in $1/S$.  We show that although  $1/S$-corrections are present
in the whole Brillouin zone (BZ), they are less drastic than in the isotropic
case of the triangular lattice.  This is due
DM interactions, which suppress
quantum fluctuations and open a gap at the ordering vector.  In
Sec.~\ref{sec:ram}, we first review the LF formalism and then use it to calculate
the Raman spectra at various levels of approximation, i.e., using only the
bare magnon dispersion, using a magnon dispersion renormalized to order
1/S, and with final state interactions included. We show that the Raman
profile is very sensitive to the magnon-magnon interactions and to the scattering
geometry. Finally, Sec.~\ref{sec:summary} presents a summary of the work.

\section{Model}\label{sec:model}
We start with the following spin-1/2 model Hamiltonian for Cs$_2$CuCl$_4$:
\begin{equation}
    \label{eq:H}
    H=\sum_{\langle ij\rangle} \bigl[
    J_{ij}\,{\bf S}_i \cdot {\bf S}_j + \mathbf D_{ij}\cdot (\mathbf
    S_i \times \mathbf S_j)\bigr],
\end{equation}
where $\langle ij \rangle$ refers to nearest-neighbor (NN) bonds on the
triangular lattice, and $J_{ij}$ and $\mathbf D_{ij}$ are the symmetric
and antisymmetric exchange constants. The
antisymmetric spin exchange originating from the relativistic spin-orbit
interaction is also known as Dzyaloshinskii-Moriya (DM) interaction.
For Cs$_2$CuCl$_4$, it is customary to
denote the exchange constants $J_{ij}$ along the horizontal bonds, which form quasi-one-dimensional chains,
as $J$, and $J_{ij}$ along the zigzag bonds as $J'$. \cite{coldea1,coldea2}
  In this paper we consider the DM vectors in the geometry suggested by the neutron-scattering work
by Coldea et al..\cite{coldea2} In this geometry
the DM interaction vanishes along the quasi-1D chains, whereas on the zigzag bonds, the
DM vectors are perpendicular to the triangular plane $\mathbf D_{ij} = \pm (0, 0, D)$ (Fig.~\ref{fig:tri}).
Experimental measurements in high magnetic field have resulted in $J\approx 0.374$ meV,
$J'\approx 0.128$ meV, and $D = 0.02$ meV.\cite{coldea2}

The classical ground state of Hamiltonian~(\ref{eq:H}) is given by a
spin spiral $\mathbf S_i/S = \cos\bigl(\mathbf Q\cdot\mathbf
r_i)\,\hat\mathbf b + \sin\bigl(\mathbf Q\cdot\mathbf r_i\bigr)\,
\hat\mathbf c$, where the pitch vector $\mathbf Q = Q \, \hat
\mathbf b$ depends on the ratio $J'/J$. In the isotropic case $J' =
J$, the ground state is the well known 120$^\circ$ non-collinear
magnetic order with $Q = 2\pi/3$.\cite{Jolicoeur1989,chubukov} To understand
 the classical ground state in detail, we consider the energy of the
magnetic spiral:
\begin{eqnarray}
    E_0(\mathbf Q) = 3N S^2 J^T_{\mathbf Q}, \quad\quad J^T_{\mathbf
    Q} = J_{\mathbf Q} - D_{\mathbf Q},
\end{eqnarray}
where
\begin{eqnarray}
    J_{\mathbf k} &=& \frac{1}{3}\Bigl(J \cos k_y + 2 J'
    \cos\frac{k_y}{2}\cos\frac{\sqrt{3} k_z}{2}\Bigr), \\
    D_{\mathbf k} &=&
    \frac{2D}{3}\sin\frac{k_y}{2}\cos\frac{\sqrt{3}k_z}{2}.
\end{eqnarray}
Minimization with respect to $Q$ leads to the following three types of
long-range magnetic order: (1) At $J'>2 J$, the magnetic spiral reduces to
 collinear N\'eel order with ferromagnetic ordering of the spins
along the chains. Along the $c$ direction, spins on  adjacent
chains are antiparallel to each other.  (2) At $0< J' \leq 2 J$, the pitch of
the spiral is given by $Q=2\arccos (-\frac{J'}{2J})$, which varies from $2\pi
\rightarrow \pi$ as we vary $J'$ from $2J$ to 0. (3) For $J'=0$, the system
degenerates into decoupled antiferromagnetic chains.

\section{Large-$S$ expansion}\label{sec:largeS}

The large-$S$ expansion about the classical spiral  order can be
significantly simplified with a locally rotated frame of reference.
\cite{chubukov} The spin components $S_i$ in a laboratory frame
are related to those in the rotated local frame through
\begin{eqnarray}
    \label{eq:spiral}
    S^x_i &=& \tilde S^x_i,  \nonumber \\
    S^y_i &=& \tilde S^y_i \cos  Q  - \tilde S^z_i \sin Q, \\
    S^z_i &=& \tilde S^y_i \sin Q  + \tilde S^z_i \cos Q
    \nonumber
\end{eqnarray}
The spiral  viewed from the rotated local frame corresponds to a simple
ferromagnetic order $\tilde \mathbf S_i = S \hat\mathbf z$. We then
employ the Holstein-Primakoff transformation:
\begin{eqnarray}
    \label{eq:HP}
    \tilde{S}^z_i &=& S - a^+_ia^{\phantom{+}}_i \\ \nonumber
    \tilde{S}^+_i &=& (2S - a^+_ia^{\phantom{+}}_i)^{1/2}\,
    a^{\phantom{+}}_i \\ \nonumber
    \tilde{S}^-_i &=& a^{+}_i (2S -a^+_ia^{\phantom{+}}_i)^{1/2}~.
\end{eqnarray}
The magnon operators  $a^{\dagger}_i$ and $a_i$ describe excitations
around the spiral ground state.  As we intend to study magnon
interactions to first order in $1/S$, we need to expand the
Hamiltonian in Eq.~(\ref{eq:H}) up to quartic order in the boson 
operators:
\begin{eqnarray}
    \label{eq:Hb}
    H = E_0+3JS(H_2+H_3+H_4).
\end{eqnarray}
Introducing Fourier transform $a_i = \sum_{\bf k} a_{{\bf k}}\, e^{i{\bf
k}\cdot{\bf r}_i}/\sqrt{N}$, the explicit expression of 
the various terms in the Hamiltonian reads
\begin{eqnarray}
    \label{eq:H2}
    H_2 = \sum_{\mathbf k} \Bigl[ A_{\mathbf
    k}\,a^{\dagger}_{\mathbf k}\,a_{\mathbf k} + \frac{B_{\mathbf
    k}}{2}\bigl(a_{\mathbf k}\,a_{-\mathbf k} + a^{\dagger}_{\mathbf
    k}\,a^{\dagger}_{-\mathbf k}\bigr)\Bigr],
\end{eqnarray}
\begin{eqnarray}
    \label{eq:H3}
    H_3 = \frac{i}{2}\sqrt{\frac{1}{2NS}} \sum_{\{\mathbf
    k_i\}}(C_{\mathbf 1} + C_{\mathbf
    2})\,\bigl(a^{\dagger}_{\mathbf 3} a_{\mathbf 1}
    a_{\mathbf 2} - a^{\dagger}_{\mathbf 1} a^{\dagger}_{\mathbf
    2} a_{\mathbf 3}\bigr),
\end{eqnarray}
\begin{eqnarray}
    \label{eq:H4}
    H_4 = \frac{1}{ 8 NS}\sum_{\{\mathbf
    k_i\}} \biggl\{\Bigl[(A_{\mathbf 1 -
    \mathbf 3} + A_{\mathbf 1 - \mathbf 4} + A_{\mathbf 2 -
    \mathbf 3} + A_{\mathbf 2-\mathbf 4}) \nonumber \\
    - (B_{\mathbf 1 -
    \mathbf 3} + B_{\mathbf 1 - \mathbf 4} + B_{\mathbf 2 -
    \mathbf 3} + B_{\mathbf 2-\mathbf 4}) \nonumber \\
    - (A_{\mathbf 1} +
    A_{\mathbf 2} + A_{\mathbf 3}
    + A_{\mathbf 4})\Bigr] a^{\dagger}_{\mathbf 1} a^{\dagger}_{\mathbf
    2} a_{\mathbf 3} a_{\mathbf 4} \nonumber \\
    - \frac{2}{3} \Bigl(B_{\mathbf 1} + B_{\mathbf 2} +
    B_{\mathbf 3}\Bigr)\,\Bigl(a^{\dagger}_{\mathbf 1}
    a^{\dagger}_{\mathbf 2} a^{\dagger}_{\mathbf 3} a_{\mathbf
    4} + a^{\dagger}_{\mathbf 4} a_{\mathbf 1} a_{\mathbf
    2} a_{\mathbf 3}\Bigr)\biggr\}.
\end{eqnarray}
Here $\mathbf 1\cdots \mathbf 4$ denote $\mathbf k_1 \cdots \mathbf
k_4$, and the summation in $H_3$ and $H_4$ is subject to momentum
conservation module a reciprocal lattice vector: $\sum_i \mathbf k_i
= 0$ mod $\mathbf G$. The following functions are introduced:
\begin{eqnarray}\label{AB}
    \begin{array}{c}
    A_{\mathbf k} = J_{\mathbf k} +
    \frac{1}{2}\bigl(J^T_{\mathbf Q + \mathbf k}
    + J^T_{\mathbf Q - \mathbf k}\bigr)
    -2J^T_{\mathbf Q},  \\   \\
    B_{\mathbf k} = \frac{1}{2}\bigl(J^T_{\mathbf Q + \mathbf k}
    + J^T_{\mathbf Q - \mathbf k}\bigr)-J_{\mathbf k}, \quad\quad
    C_{\mathbf k} = J^T_{\mathbf Q + \mathbf k} - J^T_{\mathbf Q -
    \mathbf k}.
    \end{array}
\end{eqnarray}
Here, for convenience, we rescale  the interactions $J_{\mathbf k}$ and $J^T_{\mathbf k}$ with respect to $J$, which
 is assumed to be $J=1$.

*
\begin{figure}
\includegraphics[width=0.8\columnwidth]{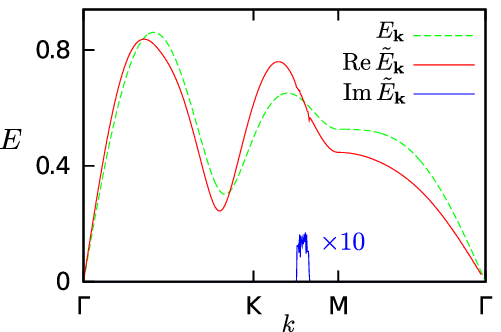}
\vspace{0.2cm}
\includegraphics[width=0.9\columnwidth]{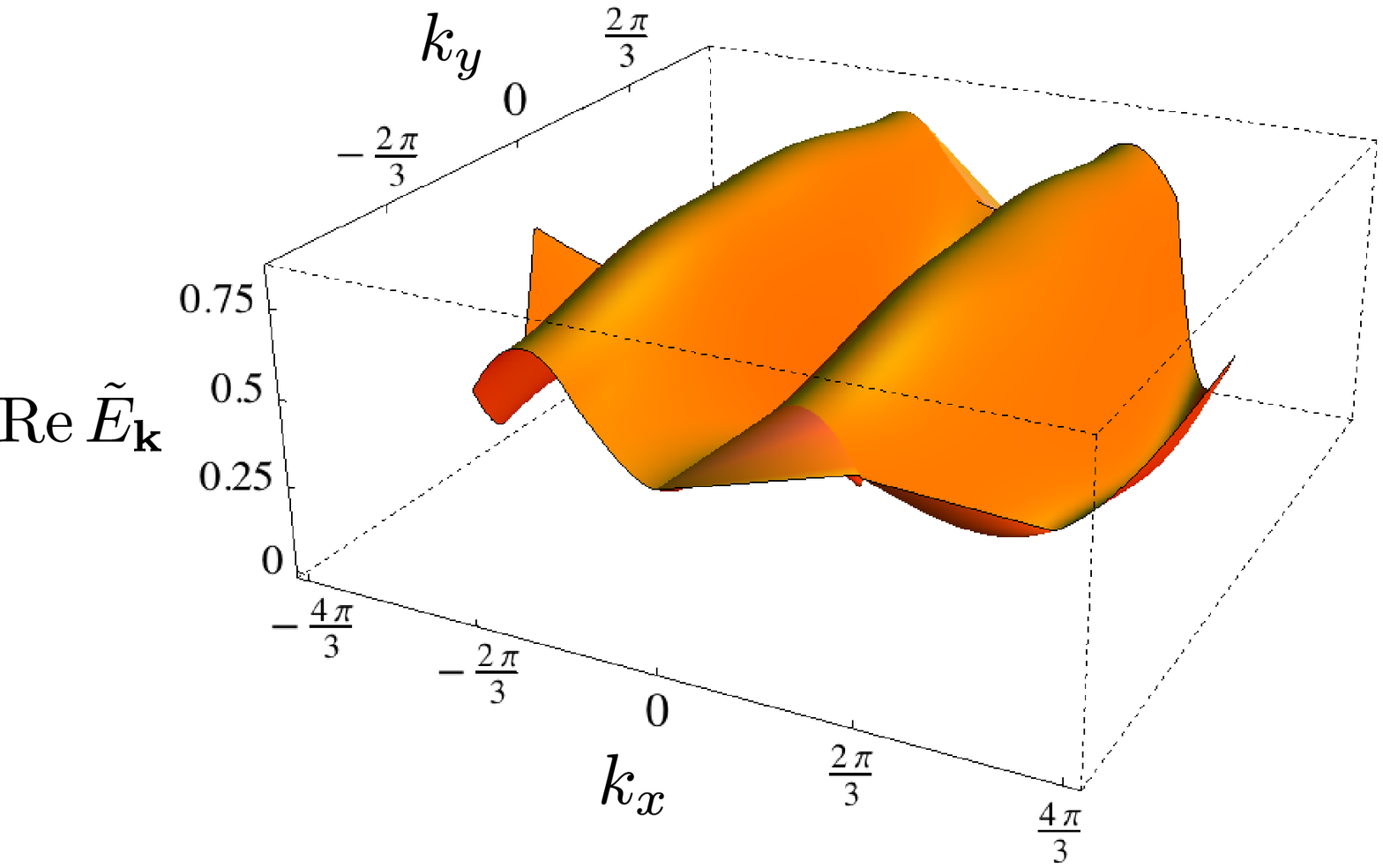}
\vspace{0.2cm}
\includegraphics[width=0.9\columnwidth]{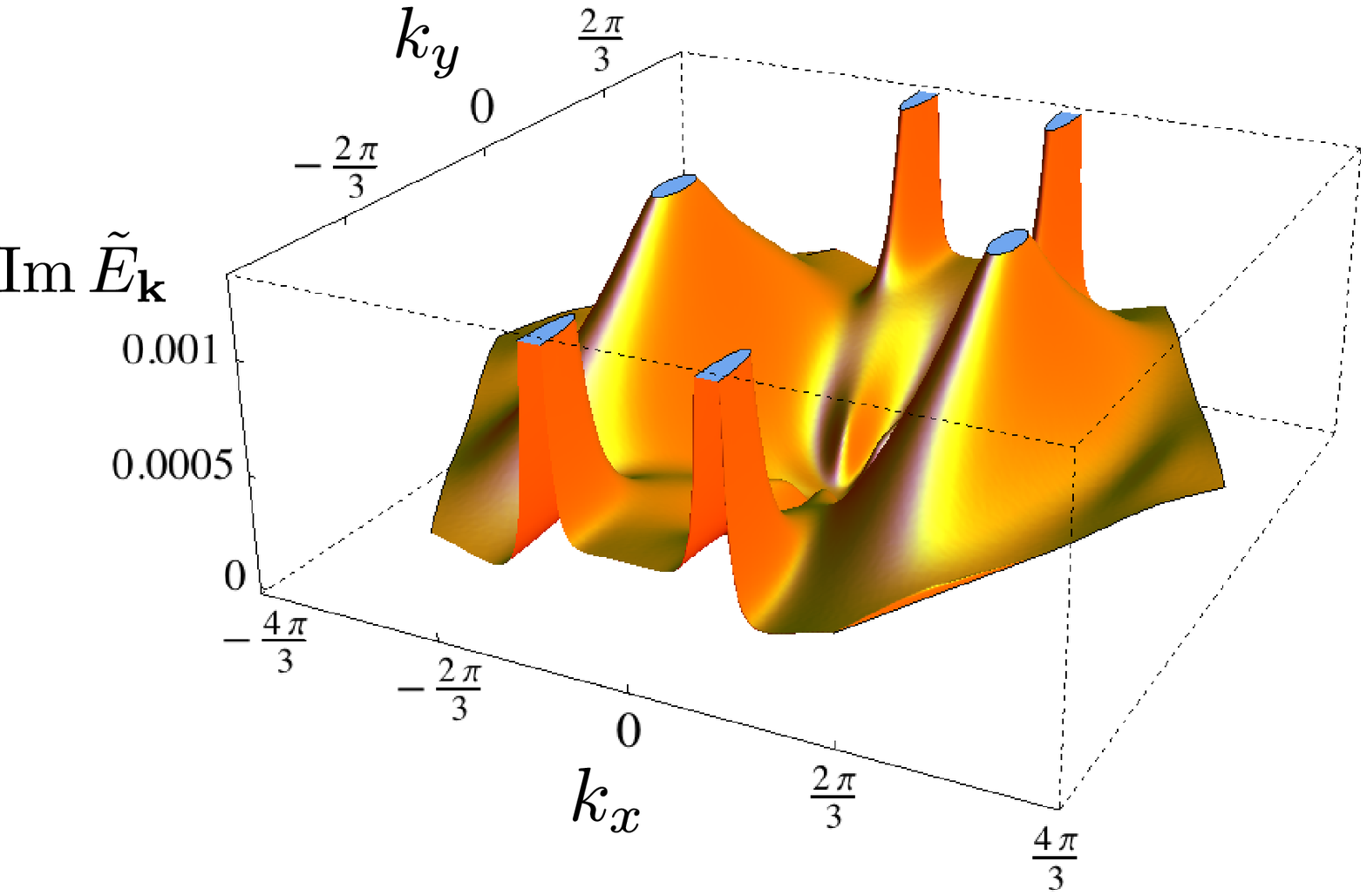}
\caption{\label{disp} Top: Renormalized magnon dispersion. Dotted line
  corresponds to the linear spin-wave dispersion $E_{\bf k}$, while red and blue
  solid lines correspond to the real and imaginary part $Re(Im) {\tilde E}_{\bf
    k}$, respectively, computed on a lattice of $252\times 252$ ${\bf k}$-points
  with artificial line broadening of $\eta =0.003$. Middle and Bottom: 3D-plot
  of the $Re{\tilde E}_{\bf k}$ and of the $Im {\tilde E}_{\bf k}$,
  respectively.  The spectrum is computed for $J=1$, while other parameters
  describing interactions is Cs$_2$CuCl$_4$, are correspondingly rescaled. All
  energies are measured in units of $3JS$.}
\end{figure}

The quadratic Hamiltonian $H_2$ can then be diagonalized by a
Bogoliubov transformation:
\begin{eqnarray}
    \label{bt} a^{\phantom{\dagger}}_{\bf k}&=& u_{\bf
    k}c^{\phantom{\dagger}}_{\bf k}+ v_{\bf k}c_{-{\bf k}}^{\dagger}
    \\\nonumber
    a_{\bf k}^{\dagger}&=& u_{\bf k}c_{\bf k}^{\dagger}+ v_{\bf
    k}c^{\phantom{\dagger}}_{-{\bf k}}~,
\end{eqnarray}
where $c^{(\dagger)}_{\bf k}$ are operators for Bogoliubov quasiparticles. The coherence coefficients
\begin{eqnarray}
    \label{uv} u_{\bf k} = \sqrt{\frac{A_{\bf k}+E_{\bf k}}{2E_{\bf
    k}}}, \quad\quad v_{\bf k}= -\frac{B_{\bf k}}{|B_{\bf k}|}
    \sqrt{\frac{A_{\bf k}- E_{\bf k}}{2E_{\bf k}}}~
\end{eqnarray}
satisfy $u_{\bf k}^2-v_{\bf k}^2=1$, and
\begin{eqnarray}\label{en}
E_{\bf k}=\sqrt{A_{\bf k}^2-B_{\bf k}^2}
\end{eqnarray}
describes the quasiparticle dispersion (the
energy of the quasiparticles is $3JS E_{\mathbf k}$). The
diagonalized Hamiltonian $H_2$ is given by
\begin{eqnarray}
    \label{h2diag}
    H_2 = E_2({\bf Q})+ \sum_{\bf k}
    E_{\bf k}c_{\bf k}^{\dagger}c_{\bf k}~,
\end{eqnarray}
where
\begin{eqnarray}
    \label{e2}
    E_2({\bf Q})&=& \sum_{\mathbf k} \bigl( A_{\mathbf k}
    v_{\bf k}^2 + B_{\mathbf k} u_{\bf k}v_{\bf k}\bigr) \nonumber
    \\
    &=& -N
    J^T_{\mathbf Q} + \frac{1}{2} \sum_{\mathbf k} E_{\mathbf
    k}.
\end{eqnarray}
gives $1/S$ correction to the classical ground state energy
$E_0({\bf Q})$.
The $1/S$ correction to the ordering wave vector $\mathbf Q$
is determined by minimizing the sum
\begin{eqnarray}
    E_0 + E_2 = 3JS(S-1) N J^T_{\mathbf Q} + \frac{3 J S}{2}
    \sum_{\mathbf k} E_{\mathbf k}
\end{eqnarray}
with respect to $Q$. The quantum correction is given by
\begin{eqnarray}\label{deltaQ}
    \Delta Q =  \frac{-1}{\partial^2 J^T_Q/\partial
    Q^2}\,\frac{1}{N}\sum_{\mathbf k} \frac{A_{\mathbf k} -
    B_{\mathbf k}}{2 E_{\mathbf k}}\,\frac{\partial J^T_{\mathbf Q +
    \mathbf k}}{\partial Q}\bigg|_{Q_0},
\end{eqnarray}
where $Q_0 = 2\arccos (-\frac{J'}{2J})$ is the pitch of the
classical ground state.

The $1/S$ contribution from the quartic Hamiltonian can be obtained
through a mean-field decoupling of  $H_4$. We first define
\begin{eqnarray}
    G(\mathbf k) = \langle a^{\dagger}_{\mathbf k}\,a_{\mathbf
    k}
    \rangle, \quad
    F(\mathbf k) = \langle a_{\mathbf k}\,a_{-\mathbf k}
    \rangle = \langle a^{\dagger}_{\mathbf
    k}\,a^{\dagger}_{-\mathbf k} \rangle.
\end{eqnarray}
The quadratic Hamiltonian plus the decoupled $H_4$ can be
expressed as:
\begin{equation}\label{h4deca}
    H_2 + {\bar H}_4
    =  3JS \sum_{\mathbf k} \Bigl[ \bar A_{\mathbf
    k}\,a^{\dagger}_{\mathbf k}\,a_{\mathbf k} + \frac{\bar B_{{\mathbf
    k}}}{2}\,\bigl(a_{\mathbf k}\,a_{-\mathbf k} + a^{\dagger}_{\mathbf
    k}\,a^{\dagger}_{-\mathbf k}\bigr)\Bigr],
\end{equation}
where
\begin{eqnarray}\label{a4b4}
    \bar A_{\mathbf k} &=& A_{\mathbf k} +
    \frac{1}{N S} \sum_{\mathbf q} \Bigl[\Bigl(A_{\mathbf k
    -\mathbf q} + B_{\mathbf k - \mathbf q} \nonumber \\
    & & \,\,- A_{\mathbf k} - A_{\mathbf q}\Bigr)\,G(\mathbf q)
    - \Bigl(B_{\mathbf q} + \frac{B_{\mathbf
    k}}{2}\Bigr)\,F(\mathbf q)\Bigr],
\end{eqnarray}
\begin{eqnarray}
    \bar B_{\mathbf k} &=& B_{\mathbf k} + \frac{1}{N S} \sum_{\mathbf q}
    \Bigl[-\Bigl(B_{\mathbf k} + \frac{B_{\mathbf q}}{2}\Bigr)\,G(\mathbf q)
    \nonumber \\ & & + \Bigl(A_{\mathbf k-\mathbf q}
    + B_{\mathbf k - \mathbf q} -
    \frac{A_{\mathbf q}}{2} - \frac{A_{\mathbf k}}{2}\Bigr)\,F(\mathbf q)\Bigr].
\end{eqnarray}
The magnon spectrum renormalized by the quartic Hamiltonian $H_4$ becomes
\begin{eqnarray}\label{h4cor}
{\bar E}_{\bf k}=\sqrt{\bar A_{\mathbf k}^2 - \bar B_{\mathbf k}^2}
= E_{\bf k}+\Sigma^{(4)}_{\mathbf k} + \mathcal{O}(1/S^2)~,
\end{eqnarray}
where $\Sigma^{(4)}(\mathbf k)$ is self-energy correction of order
$1/S$:
\begin{eqnarray}
    \Sigma^{(4)}_{\mathbf k} &=&
    A^{(4)}_{\bf k}(u^2_{\mathbf k}+ v^2_{\mathbf k}) +2B^{(4)}_{\bf k}u_{\mathbf k}v_{\mathbf
    k} \nonumber \\
    &=& \bigl(A_{\mathbf k}\,A^{(4)}_{\mathbf k}
    - B_{\mathbf k}\,B^{(4)}_{\mathbf k}\bigr)/E_{\mathbf k}.
\end{eqnarray}

To obtain the $1/S$ correction from the cubic Hamiltonian $H_3$, we
follow Ref.~\onlinecite{chubukov} and consider interactions between
quasiparticles $c$, $c^\dagger$:
\begin{eqnarray}\label{h3final}
    H_3 &=& \frac{i}{4}\sqrt{\frac{1}{2 N S}}\sum_{\{\mathbf
    k_i\}}\Bigl[\Phi_1(\mathbf k_1, \mathbf k_2;\mathbf
    k_3)\,c^{\dagger}_{\mathbf k_1} c^{\dagger}_{\mathbf k_2}
    c_{\mathbf k_3} \nonumber \\
    & & + \frac{1}{3}\Phi_2(\mathbf k_1, \mathbf k_2,
    \mathbf k_3)\,c^{\dagger}_{\mathbf k_1} c^{\dagger}_{\mathbf k_2}
    c^{\dagger}_{\mathbf k_3}\Bigr] + \mbox{h.c.}
\end{eqnarray}
The vertex functions are given by (for simplicity, we denote $1
\equiv \mathbf k_1$, $2 \equiv \mathbf k_2$, etc.)
\begin{eqnarray}\label{f1}
    \Phi_1(1,2;3) =
    \frac{\tilde\Phi_1(1,2;3)}{\sqrt{E_1 E_2 E_3}},
    \quad     \Phi_2(1,2,3) =
    \frac{\tilde\Phi_2(1,2,3)}{\sqrt{E_1 E_2 E_3}}.
\end{eqnarray}
where
\begin{eqnarray}\label{f2}
    \tilde\Phi_1(1,2;3) = C_1 f^{(1)}_-(f^{(2)}_+ f^{(3)}_+ + f^{(2)}_-
    f^{(3)}_-) \nonumber \\
    + C_2 f^{(2)}_-(f^{(3)}_+ f^{(1)}_+ + f^{(3)}_-
    f^{(1)}_-)  \nonumber \\
    + C_3 f^{(3)}_-(f^{(1)}_+ f^{(2)}_+ - f^{(1)}_-
    f^{(2)}_-)
\end{eqnarray}
\begin{eqnarray}
    \tilde\Phi_2(1,2,3) = C_1 f^{(1)}_-(f^{(2)}_+ f^{(3)}_+ -
    f^{(2)}_- f^{(3)}_-) \nonumber \\
    + C_2 f^{(2)}_-(f^{(3)}_+ f^{(1)}_+ - f^{(3)}_-
    f^{(1)}_-) \nonumber \\
    + C_3 f^{(3)}_-(f^{(1)}_+ f^{(2)}_+ - f^{(1)}_-
    f^{(2)}_-)
\end{eqnarray}
 and $f^{(\alpha)}_{\pm}=\sqrt{A_{\alpha}\pm B_{\alpha}}$ for
 $\alpha=1,2,3$.

The triplic contribution to the self-energy is
\begin{widetext}
\begin{eqnarray}\label{sigma3}
    \Sigma^{(3)}_{\mathbf k} =
    -\frac{1}{16 N S}
    \Biggl( \sum_{\mathbf k_1
    + \mathbf k_2 = \mathbf k} \frac{|\Phi^{(1)}(\mathbf k_1, \mathbf k_2, \mathbf
    k)|^2}{E_{\mathbf k_1} +
    E_{\mathbf k_2} - E_{\mathbf
    k} + i\,\eta} + \sum_{\mathbf k_1
    + \mathbf k_2 = - \mathbf k} \frac{|\Phi^{(2)}(\mathbf k_1, \mathbf k_2, \mathbf
    k)|^2}{E_{\mathbf k_1} +
    E_{\mathbf k_2} + E_{\mathbf
    k} + i\,\eta}\Biggr).
\end{eqnarray}
\end{widetext}
The first term in Eq.(\ref{sigma3}) describes a virtual decay of a
magnon into two-particle intermediate states. Terms with three
creation(annihilation) operators, as will become clear in the next
section, play no role in evaluating the magnon interactions within
the Raman response. We also note that  $\Sigma^{(3)}_{\mathbf k}$  is computed within so-called on-shell approximation.
In this approximation the self-energy is evaluated at the bare magnon energy $E_{\mathbf k}$.

Finally, the magnon energy renormalized by  both the quartic and the triplic terms  is given by
\begin{eqnarray}\label{renorm}
{\tilde E}_{\mathbf k}
= E_{\mathbf k}+\Sigma^{(4)}_{\mathbf k} +  \Sigma^{(3)}_{\mathbf k}+ \mathcal{O}(1/S^2)~.
\end{eqnarray}

In Fig.~\ref{disp} we
plot
the renormalized magnon spectrum ${\tilde E}_{\mathbf k}$ for parameters relevant to Cs$_2$CuCl$_4$.  One can see
that the renormalization of the spectrum is much less pronounced than in the
case of the isotropic triangular lattice.\cite{chubukov06,mike} Moreover, the
imaginary part of the magnon energy, $Im{\tilde E}_{\mathbf k}$ almost
vanishes in the whole BZ.  Thus, the life time of the quasi-particles is
  very large at almost any momentum. This happens because the DM interaction
opens a gap at the ordering $\mathbf Q$ vector, which significantly
suppresses quantum fluctuations.\cite{dalidovich06} We note that, as corrections
to the ordering vector $\mathbf Q$, determined by
Eq.(\ref{deltaQ}), are small at those values of $D$ relevant to our
  calculation, we  use the bare value of the
ordering wave vector ${\mathbf Q}_0$ for the remainder of the paper.

\section{Raman intensity}\label{sec:ram}

\subsection{Loudon-Fleury formalism}

Here we  present the  analysis of the two-magnon Raman scattering  from the anisotropic triangular lattice. We  employ
the LF approach which models the interaction of light
with spin degrees of freedom.
The LF scattering operator  is given by  the photon-induced super-exchange operator\cite{fleury,comment}
\begin{eqnarray}
    R=\sum_{i,\pm\bm\delta_{\mu}}(\hat{\bm\varepsilon}_{\rm
    in}\cdot {\bm\delta}_{\mu}) (\hat{\bm\varepsilon}_{\rm out}\cdot
    {\bm\delta}_{\mu})\, J_{\mu}\,{\bf S}_i\cdot{\bf S}_{i\pm\bm\delta_{\mu}}~, \label{loudonf}
\end{eqnarray}
weere $\bm\delta_{\mu}$ denote basic vectors of triangular lattice: $\bm\delta_{1}=(1,0)$,  $\bm\delta_{2}=(\frac{1}{2},\frac{\sqrt{3}}{2})$
 and $\bm\delta_{3}=(-\frac{1}{2},\frac{\sqrt{3}}{2})$. $J_{\mu}$ defines the  Heisenberg exchange on the bond ${\bm\delta}_{\mu}$.
Since $J_{\mu}$ is  anisotropic, the $C_{3v}$ symmetry of the triangular lattice is broken. Thus, instead of using $C_{3v}$-irreducible
representations ($A_1$, $A_2$ and
$E$) for characterization of the polarizations, we   determine polarizations of  incoming  and  outgoing light as
$\hat{\bm\varepsilon}_{\rm in}=\cos\theta \,\hat{\mathbf x}+\sin\theta \,\hat {\mathbf y}$
and $\hat{\bm\varepsilon}_{\rm out}=\cos\phi \,\hat {\mathbf x}+\sin\phi\, \hat {\mathbf y}$,
where $\theta$ and $\phi$  are defined with respect to the $x$-axis.

In terms of Bogoliubov quasi-particle $c$-operators, the LF scattering  operator (\ref{loudonf}) takes the
following form:
\begin{eqnarray}
    R=\sum_{\mathbf k} M_{\mathbf k}(c_{\mathbf k}c_{-{\mathbf k}}+ c_{\mathbf
    k}^{\dagger}c_{-{\mathbf k}}^{\dagger}) \equiv r^- + r^+ ~, \label{lf2}
\end{eqnarray}
where $M_{\mathbf k}$ is bare Raman vertex, which is determined 
by the the magnon spectrum and by scattering geometry. The  expression for $M_{\mathbf k}$ is given by
\begin{eqnarray}
M_{\mathbf k}=F_1({\mathbf k},\theta,\phi)u_{\mathbf
k}v_{\mathbf k}+F_2({\mathbf k},\theta,\phi)(u_{\mathbf
k}^2+v_{\mathbf k}^2)~,
\end{eqnarray}
where we introduced the following notations:
 \begin{eqnarray}
\begin{array}{ll}
&F_1({\mathbf k},\theta,\phi)=2S\sum_{\mu=1}^3f_{\mu}(\theta,\phi)\xi_{{\mu}{\mathbf k}}~, \\[.2cm]
&F_2({\mathbf k},\theta,\phi)=S\sum_{\mu=1}^3f_{\mu}(\theta,\phi)\nu_{{\mu}{\mathbf k}}~,
\end{array}
\label{fk1fk2}
\end{eqnarray}
and
 \begin{eqnarray}
\xi_{1{\mathbf k}}&=&
 \cos k_x\bigl(1+\cos Q_0\bigr)-2\cos Q_0 , \nonumber \\
\xi_{2{\mathbf k}}&=&
\cos \Bigl(\frac{k_x}{2}+\frac{\sqrt{3} k_y}{2}\Bigr)\Bigl(1+\cos \frac{Q_0}{2}\Bigr) -2\cos \frac{Q_0}{2} , \nonumber\\
\xi_{3{\mathbf k}}&=&
 \cos \Bigl(\frac{k_x}{2}-\frac{\sqrt{3} k_y}{2}\Bigr)\Bigl(1+\cos \frac{Q_0}{2}\Bigr)-2\cos \frac{Q_0}{2} , \nonumber\\
\nu_{1{\mathbf k}}&=&\cos k_x \bigl( 1-\cos Q_0 \bigr) , \\
\nu_{2{\mathbf k}}&=&\cos \Bigl(\frac{k_x}{2}+\frac{\sqrt{3} k_y}{2}\Bigr)
\Bigl(1-\cos \frac{Q_0}{2} \Bigr) , \nonumber\\
\nu_{3{\mathbf k}}&=&\cos \Bigl(\frac{k_x}{2}-\frac{\sqrt{3} k_y}{2}\Bigr)
\Bigl( 1-\cos \frac{Q_0}{2}\Bigr) . \nonumber
\label{xinu}
\end{eqnarray}
The functions $f_{\mu}(\theta ,\phi) \equiv (\hat{\bm\varepsilon}_{\rm in}\cdot
{\bm\delta}_{\mu})(\hat{\bm\varepsilon}_{\rm out}\cdot
{\bm\delta}_{\mu})$ are symmetry weighting factors  along the three
 basic vectors $\bm \delta_\mu$ of the  triangular lattice.

The Raman intensity  is obtained from the Fermi's golden rule,
which on the imaginary frequency axis reads
\begin{equation}
I(\omega_m) \simeq\Im {\rm m} \bigl[ \int_0^\beta d\tau\,
e^{i \omega_m \tau} \left\langle T_\tau ( R(\tau) R ) \right\rangle \bigr]~. \label{Fermi}
\end{equation}
By analytically continuing Matsubara frequencies
$\omega_m = 2\pi m T $ onto the real axis as $i\omega_m\rightarrow \Omega+i\eta$, we obtain the
Raman intensity as a function of the inelastic energy transfer $\Omega=\omega_{\rm in}-\omega_{\rm out}$ of the incident photons.
For the rest of the paper we assume the temperature $T=1/\beta$ to be zero.

To order $1/S$, the Raman polarization operator $\left\langle T_\tau(
R(\tau) R) \right\rangle$ contains only terms like $\left\langle T_\tau(r^{+}(\tau) r^-)\right\rangle
+\left\langle T_\tau (r^-(\tau) r^{+})\right\rangle$, where
$r^\pm$ are specified in Eq.~(\ref{lf2}) and Fig.~\ref{gamam}a).
By Hermitian conjugation, it is sufficient to calculate $J(\tau)=\left\langle
T_\tau (r^-(\tau) r^+)\right\rangle$, which is depicted in Fig.~\ref{gamam}b).

\subsection{Raman intensity  without  final state interactions}

\begin{figure}
\includegraphics[width=0.9\columnwidth]{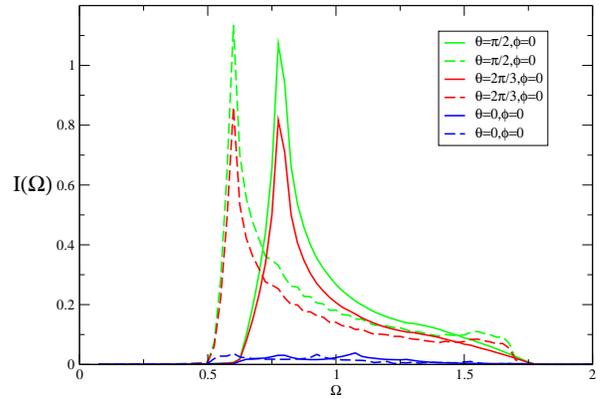}
\caption{Bare Raman intensity (solid lines) and Raman intensity including the
  one-magnon renormalization of the spectrum (dashed lines) at different
  polarizations of light described by $\phi$, and $\theta$ scattering
  angles. Number of ${\mathbf k}$-points: $252\times 252$.  The
  broadening parameter is $\eta =0.003$. $\Omega$ is measured in units of $3JS$.} \label{intensity}
\vspace{0.5cm}
\end{figure}

We now focus on the contribution from two-magnon Raman scattering, which
can be computed at different levels of approximation.
We begin by first using only the bare spin-wave dispersion. Then we continue by including
renormalizations of the one-magnon spectrum to $1/S$ order.
Calculating the Raman polarization operator $\left\langle T_\tau( R(\tau) R)
\right\rangle$ with the bare propagators of the Bogoliubov
quasiparticles, $G({\mathbf k}, i\omega_n)=1/(i\omega_n- E_{\mathbf k})$, where
$E_{\mathbf k}$ is the bare magnon spectrum gives the following expression for the Raman intensity:
\begin{eqnarray}\label{iomega}
I(\Omega)&\simeq& \Im {\rm m}\, \Bigl [
\int d^2{\mathbf k}\int d\omega M_{\mathbf k}^2
\frac{1}{\imath\omega-E_{\mathbf k}}\frac{1}{\imath(\Omega-\omega)-E_{\bf k}}\Bigr ]\nonumber\\
&=&-2\pi\,\Im {\rm m}\, \Bigl [\,\int d^2{\mathbf k}\, M_{\mathbf k}^2\, \frac{1}{\Omega-2E_{\mathbf k}+\imath\eta}\Bigr ] \\\nonumber
&=&2\pi\eta \int d^2{\mathbf k}\, M_{\mathbf k}^2\, \frac{1}{(\Omega-2E_{\mathbf k})^2+\eta^2}
\end{eqnarray}

Fig.~\ref{intensity} shows the bare Raman spectra (solid lines) as functions
of the transferred photon frequencies $\Omega$ for three scattering geometries:
(i) $\theta=\pi/2,~\phi=0$, (ii) $\theta=2\pi/3,~\phi=0$ and (iii) $\theta=0,~\phi=0$.
For two of the polarizations (i) and (ii),
the Raman spectra show a similar profile and intensity: the Raman response exhibits a peak at $\Omega\simeq 0.8-0.9$,
and the location of this peak corresponds to the twice the energy of the dominant
van-Hove singularity of the one-magnon dispersion, cf. (Fig.~\ref{disp}).

\begin{figure}
\includegraphics[width=0.8\columnwidth]{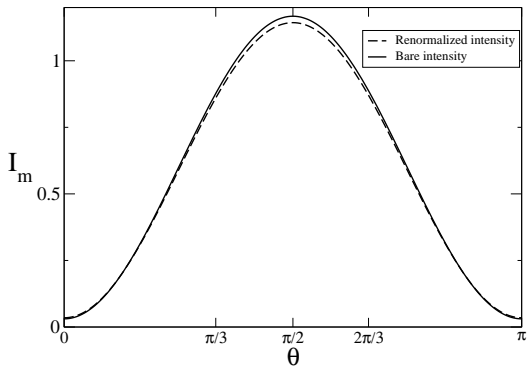}
\caption{Variation of the maximum  of the  bare  and renormalized intensities as a function of the scattering angle $\theta$ computed at $\Omega=0.8$ and $\Omega=0.6$, respectively.} \label{thetadep}
\end{figure}

The small Raman intensity $I(\Omega)$ at low energies is due to two reasons.
First, the magnon energy is gapless only at the zone center and has a gap caused by the
DM interaction at the ordering wave vector ${\mathbf Q}_0$. Thus, only magnons
with momentum near ${\bf k}\simeq 0$ can be excited by photons with frequency
$\Omega < 2\,E( {\mathbf Q}_0)$.  As one can see from Fig.~\ref{disp}, the rather steep
magnon spectrum in the vicinity of zone center gives rise to a  small density of states.
Second, the form of $M_{\mathbf k}^2$  is such that it selects mostly wavevectors $k\sim\pi$ where the gap resides.

In the $\theta=\phi=0$ geometry, the Raman response is non-vanishing but very
small for the whole energy range
compared with intensities observed in other geometries.  The relative smallness can be understood by comparing
this result with the
well-studied case of the square lattice, for which the LF operator in the
$A_{1g}$ geometry commutes with the Heisenberg Hamiltonian, and, as a result,
the Raman response vanishes.  In the case of the anisotropic triangular lattice,
a non-commuting part of the LF operator (\ref{loudonf}) remains non-zero even for the A$_{1g}$
geometry. This  part leads to small Raman intensities, scaling with the ratio $(J'/J)^2$. Indeed, as
we have discussed earlier, the anisotropic triangular lattice can be viewed
as
an interpolation between the square and the isotropic triangular lattice by varying $J'/J$.

In Fig.~\ref{thetadep}, the two-magnon peak is shown as a function of the scattering angle $\theta$.
A strong dependence on the scattering geometry can be seen:
the largest peak intensity is observed at the $\theta=\pi/2,\phi=0$, i.e. cross-polarization, which gradually decreases
and reaches its minimum at $\theta=0$ or $\pi$ (Though not exactly, the peak intensity is roughly proportional to
$\sin^2\theta$).  This polarization dependence
is consistent with the fact that the anisotropy in the present case resembles
more that of a rectangular than of an isotropic triangular geometry.

An angular-dependence analysis of the Raman spectrum on Cs$_2$CuCl$_4$ is
highly desirable as it could provide a potential diagnosis of whether a long-range spin order develops in the ground state.
If an angular dependance similar to the one described above is observed, then the
ground state is likely to be an incommensurate spiral. \cite{veillette05,dalidovich06}
On the other hand, a Raman response that is independent of the scattering geometry
might indicate
a spin-liquid ground state,~\cite{cepas08} as proposed in some recent theories.~\cite{isakov05,alicea05,yunoki06,starykh07}
Here some precautions are necessary.
 Since the underlying Hamiltonian is spatially anisotropic, the spinon excitations of the proposed spin-liquid phase might inherit  the crystal anisotropy to some degree, hence also giving rise to
a polarization-dependent Raman signal. This question should be further studied but we believe that
 even if this is the case, the polarization dependence of the Raman response   will be much weaker than in the case
of spiral magnetic order.

Next we incorporate  $1/S$ corrections to the Raman spectrum.
This can be easily done by replacing the bare energy with the renormalized magnon energy $\tilde E_{\mathbf k}$ (30)
in the propagator $G({\mathbf k}, i\omega_n)=1/(i\omega_n- {\tilde E}_{\mathbf k})$.
The renormalized Raman spectra are shown by dashed lines in Fig.~\ref{intensity}.
We can see that in the renormalized spectrum the two-magnon peak appears at the energy
$\Omega\simeq 0.6$, which is slightly lower than the
peak in the bare spectrum.  This is in
contrast with the case of isotropic triangular lattice, where the peak is shifted to higher energies.
Once again it shows that the Raman spectra on the  anisotropic triangular lattice has features of the both triangular and square lattices.

\subsection{Raman intensity  with  final state interactions}

\begin{figure}
\includegraphics[width=0.7\columnwidth]{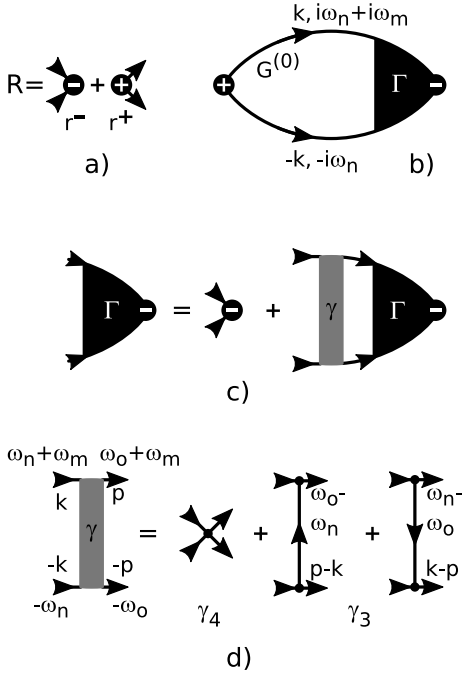}
\caption{\label{gamam}
 a) Bare Raman vertex $R$ from
Eqn. (\ref{lf2}); b) Raman susceptibility bubble; c)
The integral equation for the dressed Raman vertex $\Gamma$ in terms
of the irreducible magnon particle-particle  vertex $\gamma$;
d) Leading order $1/S$ contributions to $\gamma$.}
\end{figure}

Next we consider the final-state magnon-magnon interactions. Usually these
are not small, particularly for $S=1/2$. The effect of the final-state
magnon-magnon interactions can be taken into account by computing vertex
corrections to the bare Raman vertex.  Here we consider only the leading $1/S$ corrections.
In this approximation, the vertex corrections can be obtained from an infinite summation of ladder diagrams.
These ladder diagrams are shown in Fig.~\ref{gamam}c) in terms of the two-particle
(ir)reducible Raman vertex $(\gamma ({\mathbf k},{\mathbf p},\omega_n,\omega_o))~\Gamma({\mathbf k},\omega_n,\omega_m)$.
These are related by the Bethe-Salpeter equation:
\begin{eqnarray}\label{bseqn}
\Gamma ({\mathbf k},\omega_n,\omega_m) =
r^-({\mathbf k}) +
\sum_{{\mathbf p},\omega_o}
\gamma ({\mathbf k},{\mathbf p},\omega_n,\omega_o)
&&\nonumber\\
G({\mathbf p},\omega_o+\omega_m)
G(-{\mathbf p},-\omega_o)
\Gamma ({\mathbf p},\omega_o,\omega_m)~.&&
\end{eqnarray}
The two-particle irreducible vertex can be decomposed as in Fig.~\ref{gamam}d):
$\gamma ({\mathbf k},{\mathbf p},\omega_n,\omega_o)=\gamma_3({\mathbf
  k},{\mathbf p},\omega_n,\omega_o)+\gamma_4({\mathbf k},{\mathbf p})$.
The quartic vertex $\gamma_4({\mathbf k},{\mathbf p})$ is identical to the
two-particle-two-hole contribution from the $H_4$ term; its explicit expression is given by
\begin{widetext}
\begin{eqnarray}\nonumber
 \gamma_4({\mathbf k},{\mathbf
p})=\frac{1}{4S}
\Bigl[4\bigl(A_0 -B_0+A_{{\mathbf p}+{\mathbf k}}-B_{{\mathbf p}+{\mathbf k}}
+A_{\mathbf p}-A_{\mathbf k}\bigr)u_{\mathbf p}u_{\mathbf k}v_{\mathbf p}v_{\mathbf k}
+\\
\bigl( A_{{\mathbf p}-{\mathbf k}}-B_{{\mathbf p}-{\mathbf k}}+A_{{\mathbf p}+{\mathbf k}}-B_{{\mathbf p}+{\mathbf k}}-
A_{\mathbf p}-A_{\mathbf k}\bigr)\bigl(u_{\mathbf p}^2u_{\mathbf k}^2+v_{\mathbf p}^2v_{\mathbf k}^2\bigr)
-\\
\bigl( 2B_{{\mathbf p}}+B_{{\mathbf k}}\bigr)\bigl( u_{{\mathbf p}}^2+v_{{\mathbf p}}^2 \bigr)u_{{\mathbf k}}
v_{{\mathbf k}}-\bigl( 2B_{{\mathbf k}}+B_{{\mathbf p}}\bigr)\bigl( u_{{\mathbf k}}^2+v_{{\mathbf k}}^2 \bigr)u_{{\mathbf p}}
v_{{\mathbf p}}
\Bigr]\nonumber
\label{gamma4f}
\end{eqnarray}
\end{widetext}
where the functions $A_{\mathbf k}$ and $B_{\mathbf k}$ are  given by Eqs.~(\ref{AB}).

The triplic vertex $\gamma_3 ({\mathbf
k},{\mathbf p},\omega_n,\omega_o)$  is obtained
from the product of two vertices of the cubic term $H_3$ and one
intermediate propagator, and can be written as
\begin{widetext}
\begin{eqnarray}\label{gamma3_1}
\gamma_3 ({\mathbf k},{\mathbf  p},\omega_n,\omega_o)
=\frac{1}{32S}\,[
\Phi_1({\mathbf p}, {\mathbf k}-{\mathbf p};\mathbf
    k)\Phi^*_1({\mathbf p}, {\mathbf k}-{\mathbf p};\mathbf
    k)
G^0({\mathbf k}-{\bf p},i\omega_o-i\omega_n)
+\nonumber\\
\Phi_1({-\mathbf p}, {\mathbf p}-{\mathbf k};-\mathbf
    k)\Phi^*_1(-{\mathbf p}, {\mathbf p}-{\mathbf k};-{\mathbf k})
    G^0({\mathbf p}-{\mathbf k},i\omega_n-i\omega_o) ]~,
\end{eqnarray}
\end{widetext}
where the functions $\Phi_1(1,2;3)$ and their complex conjugates are given by
Eqs.~(\ref{f1})-(\ref{f2}).  To keep $\gamma_3({\mathbf k},{\mathbf
  p},\omega_n,\omega_o)$ to leading order in $1/S$, we retained only the zeroth
order propagators $G^0$ for each intermediate line.
We further simplify the expression (\ref{gamma3_1}) by assuming that the dominant contribution 
to the frequency summations in the Bethe-Salpeter equation (\ref{bseqn}) 
comes from the mass-shell of the  intermediate particle-particle propagators.
This corresponds to the substitution of the intermediate frequencies by
$-i\omega_n \approx E_{\mathbf k},~ -i\omega_o \approx E_{\mathbf p}~$.  The
simplified expression for the triplic vertex then reads as
\begin{eqnarray}\label{gamma3_2}
\gamma_3 ({\mathbf k},{\mathbf  p})&\simeq &\frac{1}{32S}\,\Phi_1({\mathbf p}, {\mathbf k}-{\mathbf p};\mathbf
    k)\,\Phi^*_1(-{\mathbf p}, {\mathbf p}-{\mathbf k};-\mathbf
    k)\vspace{0.2cm}\nonumber\\
     &\times&\frac{2 E_{{\mathbf k}-{\mathbf p}}} {(E_{\mathbf k}-E_{\mathbf p})^2-
    E^2_{{\mathbf k}-{\mathbf p}}}~.
\end{eqnarray}
The triplic vertex then depends only on momenta.

Next we perform the frequency summation over $\omega_o$ on the right hand side
of Eq.~(\ref{bseqn}) as well as the analytic continuation
$i\omega_m\rightarrow\Omega+i\eta\equiv z$. With this, the reducible vertex $\Gamma$ in the latter
equation turns into a function of ${\mathbf p}$ and $z$ only, leading to
\begin{eqnarray}
&&\sum_{\mathbf p} L_{{\mathbf k},{\mathbf p}}(z) \Gamma_{\mathbf
p}(z) = r^-({\mathbf k})
\label{lineqn1} \\
&&L_{{\mathbf k},{\mathbf p}}(z) = \delta_{{\mathbf k},{\mathbf p}} -
\frac{\gamma ({\mathbf k},{\mathbf p})}{z-2{\tilde E}_{\mathbf p}}~,
\label{lineqn2}
\end{eqnarray}
which is an integral equation with respect to momentum only.
Finally, the expression for the Raman intensity  can be written as
\begin{eqnarray}
I(\Omega) &\simeq& \bigl[J(\Omega)-J(-\Omega)\bigr]~,
\end{eqnarray}
where
\begin{eqnarray}
J(\Omega) &=& \Im {\rm m} \, \left[ \sum_{\mathbf k} \frac{M_{\mathbf
k}\,\Gamma _{\mathbf k}(\Omega+i\eta) }{\Omega+i\eta-2{\tilde E}_{\mathbf k}} \right]~.
\end{eqnarray}

\begin{figure}
\includegraphics[width=0.85\columnwidth]{Fig6a.eps}\\[0.5cm]
\includegraphics[width=1.0\columnwidth]{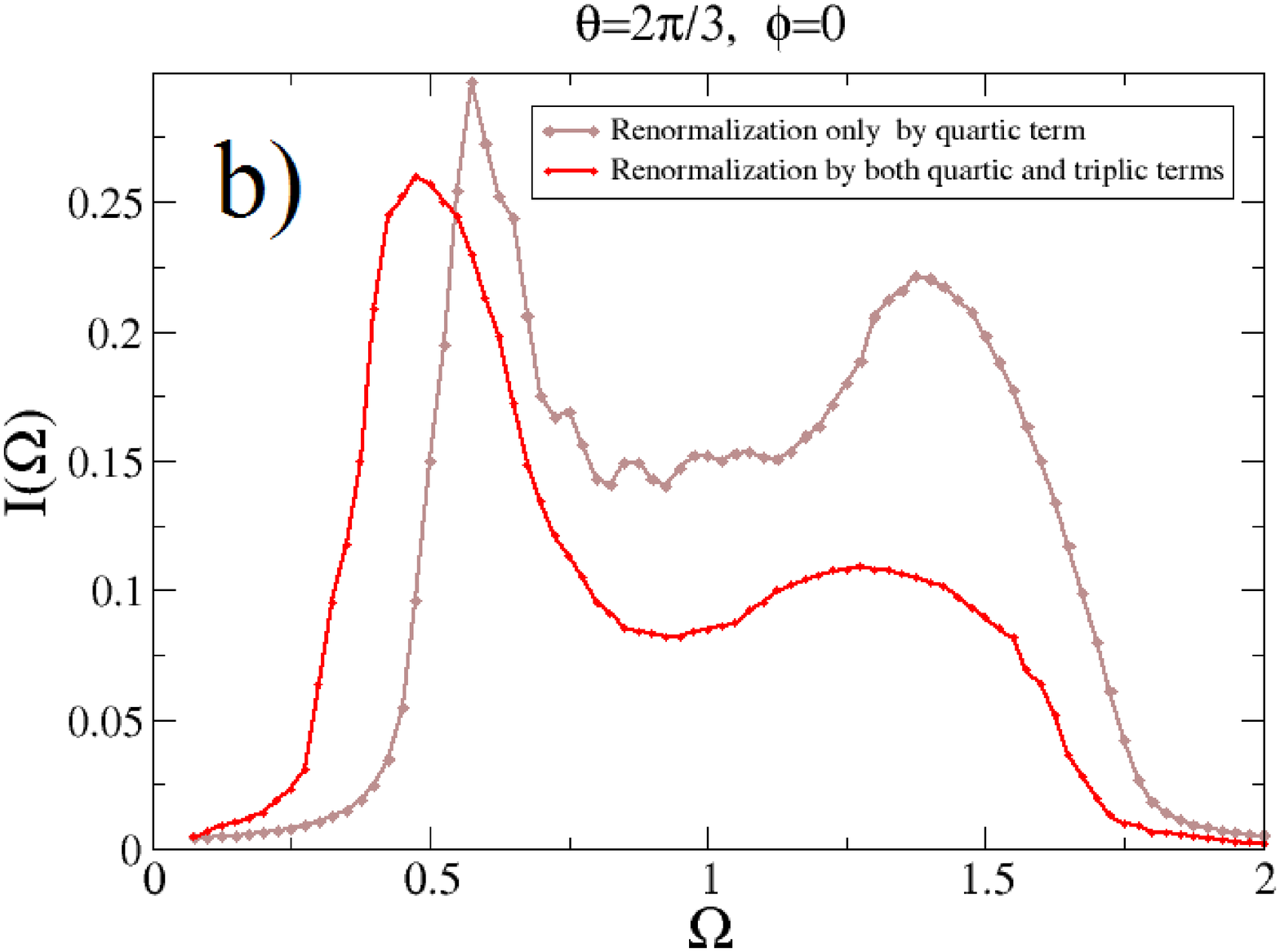}
\caption{Effect of final state interactions on Raman intensity for $\theta=2\pi/3, \phi=0$. a)
    Bare intensity, an
  intensity computed with renormalized magnon energies and intensity computed
  with included final state interactions are shown by blue, green and red
  lines, respectively.  Number of ${\mathbf k}$-points: $69\times 69$.  The
  imaginary broadening is $\eta =0.03$. $\Omega$ is measured in units of $3JS$.
 b) Comparison of the intensities computed with   corrections only due to the quartic vertex, $\gamma_4({\mathbf k})$ ( brown solid line with diamonds), and with corrections due to full vertex, $\gamma_3({\mathbf k})+\gamma_4({\mathbf k})$ ( red solid line with circles).
} \label{r}
\end{figure}

The fully renormalized intensity for the polarization with $\theta=2\pi/3,~\phi=0$ is shown in Fig.~\ref{r} a).
Despite the damping by the vertex corrections, the two-magnon peak  survives and is further shifted towards lower energies.
Apparently, the damping of the peak is less pronounced compared with the isotropic triangular lattice case.\cite{perkins08}
At the higher energies, we also see the appearance of a broad  continuum.
We would like to point out that both the peak and the broad continuum are observed in geometries
with almost perpendicular $\hat{\bm\varepsilon}_{\rm in}$ and $\hat{\bm\varepsilon}_{\rm out}$.
Although the width and intensities of the peak and the profile of the continuum
vary with the angle $\theta$, the position of the two-magnon peak and the center of the continuum do not change significantly
for $\theta \sim \pi/2$.
On the other hand, the Raman signal is extremely weak for nearly parallel geometries ($\theta$, $\phi \sim 0)$,
similar to the case without vertex corrections (see Fig.~\ref{intensity}).

In order to disentangle the contributions coming from the triplic and quartic
terms, we also compute the Raman spectrum with an irreducible vertex
which includes only the quartic part.
The comparison between the Raman spectrum computed with the full vertex and with the one containing only
$\gamma_4(\mathbf k,\mathbf p)$ is presented in Fig.~\ref{r} b). One observes that
vertex corrections due to the quartic term split the sharp two-magnon peak into
two peaks of comparable intensities, which are, however, about the half of the  intensity of the two magnon peak with final-state interactions. The triplic term modifies these two peaks quite differently.  The lower energy  peak is only weakly renormalized  by the triplic  term, while the higher energy peak is  damped more strongly  and is transformed into the broad continuum.

Finally we note that the direct comparison of these results with Fig. \ref{intensity} should be taken with a certain precaution, since the artificial line broadening in Fig.~\ref{r} is larger by one order of magnitude.
This is a consequence of a factor of 16 less $\mathbf k$-points used in the latter case.
This is because the kernel $L_{{\mathbf k},{\mathbf p}}(z)$ in the integral equation (\ref{lineqn1}) and (\ref{lineqn2})
is not sparse and has rank $N^2 \times N^2$. Consequently, a moderate lattice size gives rise
to a rather large dimension for the kernel. In the above calculations, we have chosen $N=69$,
leading to a 4761$\times$4761 system which we have solved 100 times to account for 100
frequencies in the interval $\Omega\in [0,2]$.
We also note that the kernel has points of singular behavior in $({\mathbf k},{\mathbf p})$-space,
which stem from the singularities of the Bogoliubov
factors and from the energy denominators in vertex functions.  Here, we have
chosen to regularize these points by cutting off eventual singularities in
$L_{{\mathbf k},{\mathbf p}}$.  This can be justified because the weight of
these points is negligibly small compared with the total number of points in the
BZ. We have checked that this regularization does not significantly effect the
obtained spectra.

\section{Summary}\label{sec:summary}

In summary, we have studied the two-magnon Raman scattering in the anisotropic
triangular Heisenberg antiferromagnet considering various levels of
approximation within a controlled $1/S$-expansion.
We have shown
that the Raman profile is sensitive to the magnon-magnon
interactions and to the scattering geometry.
The calculations indicate that the main effect of the magnon-magnon
interactions is on the shifting of the two-magnon
peak towards lower energies and on the formation of the
broad continuum at the higher energies.
We have also shown that through exchange and DM interactions,  the spatial
anisotropy of the lattice is
transferred to the magnon dispersion of the spiral magnetic order. This makes the
two-magnon Raman scattering anisotropic
and very sensitive to the scattering geometry when the spiral magnetic order is the ground state.
Our results on the polarization dependence of the spectrum suggest that Raman
spectroscopy might be very useful to resolve the ambiguity in the interpretation
of the neutron scattering experiments and to gain insight into the magnetic
structure of Cs$_2$CuCl$_4$.

 A final note. In this paper we  used as an input the orientation  of the DM vector extracted from neutron scattering in Ref. \cite{coldea2}. Recent electron spin resonance (ESR)  measurements by Povarov et al.~\cite{povarov11} suggested an alternative orientation of the DM vectors in Cs$_2$CuCl$_4$,  in which the strongest DM interaction is along the spin chains.
The direction of the  DM vector is important for  magnetic properties of Cs$_2$CuCl$_4$ in the  presence of the magnetic field because in this case both the  magnetic ground state and the excitation spectra  depend on the relative orientation of the external field and the  direction of the DM vector.\cite{starykh10,povarov11,smirnov12}
However, the direction of the DM vector does not affect Raman intensity, at least at a qualitative level. The reasoning is that the change of the orientation of the DM vector does not change the classical ground state of the model (\ref{eq:H}) and does not change 
qualitatively  the spectrum of  spin wave excitations, as in both geometries it's main role is to open the gap.   

{\it Acknowledgement.} N.P. acknowledges the support from NSF grant DMR-1005932. G.W.C. acknowledges the
the support of ICAM and NSF grant DMR-0844115.
N.P. also thank the hospitality of the visitors program at MPIPKS, where part of the work on this manuscript has been done.


\begin{thebibliography}{99}
\bibitem{comment} Here we note that in the presence of the spin-orbit coupling giving rise to the DM interaction,
 the effective Hamiltonian Eq.(\ref{loudonf}) describing the interaction of light with magnons
  shall  also  contain an antisymmetric part.\cite{fleury,lara06}   This term contributes only to the one-magnon Raman response which we do not describe here. This is the reason why we did not include  the antisymmetric term in Eq.(\ref{loudonf}).
\end{thebibliography}
\end{document}